\documentclass[10pt,letterpaper]{article}
\usepackage{opex3}

\begin{document}

\title{Theoretical study of the implications of causality when inferring metamaterial properties}
\author{A. Eugene DePrince III and Stephen K. Gray}
\email{gray@anl.gov}
\address{Center for Nanoscale Materials,
\ Argonne National Laboratory, Argonne, IL 60439}
\date{}

\begin{abstract}
The usual procedures to extract effective refractive indices for 
negative-index metamaterials are complicated by branching ambiguities in the 
inverse cosine function.  The existing methods to eliminate ambiguities either 
involve calculations with varying geometries which are inherently flawed, 
 as metamaterial properties depend strongly on geometry, or are more subjective
as one must inspect and select branches to yield effective parameters as a 
continuous function of frequency.  We propose that the Kramers-Kronig relations, 
which hold for negative-index materials, naturally guide the selection of the 
proper branches, and in fact predict negative refractive index where other 
extraction procedures alone might not.  The experimental realization of high 
quality negative-index materials requires a robust and reliable index retrieval
procedure; the combined extraction/Kramers-Kronig retrieval can be used to 
design optimal metamaterials in a simple, systemmatic, and general manner.
\end{abstract}




\section{Introduction}
Artificial materials known as metamaterials can exhibit a range of unique 
electromagnetic properties that are not observed in naturally occuring 
materials.  Obtaining a negative index of refraction is particularly desirable, 
as negative-index materials (NIM) have many potential applications including 
imaging beyond the diffraction limit and optical cloaking. Although NIM were 
characterized by Veselago \cite{REF:Veselago64} many years ago, it is only 
recently that we have seen a sudden burst in interest in NIM design and 
applications. Pendry \cite{REF:Pendry00} showed that perfect subwavelength 
imaging or superlensing would be possible with a NIM of refractive index 
$N=-1+0i$. The exciting idea here is that imaging would be limited not by the 
wavelength of light, but by the quality of the lense.  Double negative materials
and NIM were shortly thereafter experimentally realized in 
the work of Smith and coworkers \cite{REF:Smith00-1,REF:Smith00-2}.  In 
addition,  the functional range of NIM has been driven to visible frequencies
by Shalaev by incorporating the visible-range electromagnetic resonances of 
metallic nanoparticles into NIM design \cite{REF:Shalaev05}. In practice, 
sufficient conditions to yield a negative refractive index come in the form of 
$\rm{Re}(\epsilon)<0, \rm{Re}(\mu)<0$, where $\epsilon$ and $\mu$ 
represent the relative permittivity and permeability of the material, 
respectively.  Materials having simultaneously negative permeability and
permittivity are known as double negative materials.  Note, however, that these 
conditions are sufficient but not necessary to achieve a negative index.  
Necessary conditions require only that 
$\rm{Re}(\epsilon)\rm{Im}(\mu) + \rm{Im}(\epsilon)\rm{Re}(\mu) < 0$ 
\cite{REF:Depine04}.

Several classes of NIM have been realized experimentally with a few standard 
designs emerging that yield negative indices in a wide range of frequencies.  
Photonic crystals \cite{REF:PC1,REF:PC2,REF:PC3} have been shown to 
exhibit a negative index in the near-IR range, but their utility as imaging 
devices is limited by the internal structure of the crystal, which has 
dimensions on the order of the wavelength of light \cite{REF:PC4,REF:PC5}.  
Split-ring resonators (SRR) 
\cite{REF:SRR,REF:SRR1,REF:SRR2,REF:SRR3,REF:SRR4,REF:SRR5} 
can exhibit negative index at telecommunication wavelengths.
In general, the SRR provides a negative permeability that, when combined with
the negative permitivity of a metallic component, results in a negative real
component of the refractive index. The operating frequencies of the SRR can be 
shifted toward the visible rangle by miniaturization. The first class of NIM 
to reach the visible spectrum is 
perhaps the simplest and in that sense the most versatile. A pair of nanowires
separated by some dielectric has been shown theoretically and 
experimentally give a negative refractive index in the near IR and visible 
range \cite{REF:Shalaev05,REF:Shalaev06,REF:Shalaev06-2,REF:Shalaev06-3}. 
There also exist S-shaped variations on the wire-pair structure in the
spirit of the SRR known as S-shaped resonators \cite{REF:SShaped1,REF:SShaped2}.
The behavior of double S-shaped resonators can be mimicked by
a single pair of asymmetric parallel nanowires \cite{REF:SShaped3}, which bear 
strong resemblence to the simple metal-dielectric-metal sandwich described 
above.

Clearly there exist many possibilities for high quality NIM, and characterizing 
the refractive index theoretically can aid the design of such materials, 
circumventing the need for expensive experimental optimizations
of metamaterial geometeries.  Refractive index retreival methods based on 
scattering- or S-parameters 
\cite{REF:SRR4,REF:Soukoulis02,REF:Soukoulis05,REF:Kong04} are the most 
direct and robust methods for relating refractive index to theoretically 
and experimentally tractable parameters such as complex transmission and
reflection coefficients.  The extraction procedures work reasonably well in
most cases, but, occasionally, severe branching ambiguities muddle the
procedure and introduce an element of subjectivity into the index assignments.
In this paper, we suggest a simple and powerful method for automatically 
selecting the correct branches in the extraction procedure by invoking the
concept of causality.  The well-known Kramers-Kronig relations automatically
guide the extraction procedure and in some cases allow us to predict a structure
to be a NIM that would otherwise be difficult to characterize.

\section{Theory}
S-parameter retrieval procedures begin with the assumption 
that the metamaterial may be approximated by a homogeneous slab with a 
well-defined thickness, $D$.  The complex transmission and reflection 
coefficients are given by \cite{REF:Shalaev05}
\begin{equation}
\label{EQN:T}
t = e^{i k (D - 2 \delta)} \frac{E_t(\delta)}{E_i(-\delta)},
\end{equation}
and
\begin{equation}
\label{EQN:R}
r = e^{i k (D - 2 \delta)} \frac{E_r(-\delta)}{E_i(-\delta)},
\end{equation}
where $\delta$ is the distance from the center of the slab to the evaluation 
planes above and below the sample.  The reflected and 
transmitted fields are normalized by the incident light, $E_i$ at the 
reflection evalulation plane.  The refractive index for a homogeneous NIM in 
vacuum (or on a substrate with refractive index equal to unity) is given by
\begin{equation}
\label{EQN:REFIND}
N = \pm\frac{1}{k D}{\rm cos}^{-1} \bigg ( \frac{1-r^2+t^2}{2 t} \bigg ) + \frac{2 m \pi}{k D}.
\end{equation}
The sign in Eq. (\ref{EQN:REFIND}) is uniquely determined by requiring that 
Im(N) be positive.  However, the real portion of the inverse cosine function is 
prone to branching ambiguities, where seemingly valid solutions exist for all 
integers, $m$. The S-parameter retrieval is based in the 
assumption that the metamaterial may be treated as a homogeneous material, and 
as such, the macroscopic refractive index should be independent of the 
thickness of the NIM, D. To maximize the separation of the branches, the 
extraction procedure should be performed on as thin a slab as possible, and 
even then,  one should perform more than one analysis with differing values for
D.  The true branches should reveal themselves in their insensitivty to the 
choice of D.  Unfortunately, 
metamaterial properties are very sensitive to geometric parameters, and it is
virtually impossible to alter D in such a way as to maintain a constant 
refractive index. Choosing branches becomes a more subjective process wherein
one selects parameters as a continuous function of frequency 
\cite{REF:SRR4,REF:Soukoulis02}. We will show that this process is still 
flawed, as the solution can be skewed by specious discontinuities in the 
refractive index as a function of frequency.

The ambiguities in the extraction procedure are ultimately rooted in (and
may ultimately be circumvented by) the fact that Eq. (\ref{EQN:REFIND}) is 
insufficient to guarantee that the obtained refractive indices are causal; 
that is, do the real and imaginary components reflect the analytical properties 
of $\epsilon$ and $\mu$?  The well-known Kramers-Kronig (KK) relations 
\cite{REF:Smith72,REF:Saarinen04} guarantee that knowledge of the imaginary 
component of the refractive index over all frequencies implies knowledge of 
the real component by the integration
\begin{equation}
\label{EQN:KK_real}
n(\omega) - n_\infty = \frac{2}{\pi} {\rm P} \int_0^\infty \frac{\omega^\prime \kappa(\omega^\prime)}{{\omega^\prime}^2-\omega^2} \rm{d}\omega^\prime.
\end{equation}
Similarly, knowledge of the real component over all frequencies can be mapped
to the imaginary component:
\begin{equation}
\label{EQN:KK_imag}
\kappa(\omega) = -\frac{2 \omega}{\pi} {\rm P} \int_0^\infty \frac{n(\omega^\prime)-n_\infty}{{\omega^\prime}^2-\omega^2} \rm{d}\omega^\prime.
\end{equation}
Regardless of the extent to which branching issues complicate the 
extraction procedure, the KK relationship between
the real and imaginary components of the refractive index must hold at all
wavelengths.  The correct form of Im(N) is easy to obtain from Eq. 
(\ref{EQN:REFIND}); one simply selects the sign such that the imaginary 
component is positive. We can verify the validity of the similarly obtained 
real components by integrating Eq. (\ref{EQN:KK_real}) over frequencies that 
effectively span the range 0 to $\infty$.  
In practice, very accurate values for Re(N) may be obtained by numerically 
integrating Eq. (\ref{EQN:KK_real}) over a very wide, finite range of 
frequencies. The infinite-frequency limit of the refractive 
index, $n_\infty$, is given by the extraction procedure in the limit of very 
small  wavelength.  In the limit of $\omega \to \infty$, all branches in the 
inverse cosine function converge and all ambiguities are removed.  

It should be noted that we have used Kramers-Kronig relations concerning 
$N=n+i\kappa$. Somewhat less convenient for our purposes, but more fundamental,
KK relations exist for $\epsilon$ and $\mu$. The validity of our KK relations 
requires that $N$ be analytic in the upper half of the complex plane, 
$N \to n_\infty$ as $\omega \to \infty$, and no conductivity at $\omega=0$.  
These conditions in relation to metamaterials were discussed by Skaar 
\cite{REF:Skaar06}. In particular, the metamaterial must be a passive medium, 
with no gain.  However, it has also been pointed out that $N^2$ has the same 
analytical properties as $\epsilon$ and $\mu$ separately, and therefore a KK 
relation for $N^2$ would be appropriate in the case of gain 
\cite{REF:Stockman07}. We have verified numerically that the computed $n$ and 
$\kappa$ from our approach discussed below satisfy this other KK relation.
In future work, it may be interesting to consider the use of this relation 
directly.  

\section{Applications}
\begin{figure}[!htpb]
\caption {The unit cell of a two-dimensional NIM composed of two Au rods 
separated by SiO$_2$ (a). The real component of the refractive index (b) is 
clearly negative in the 850-900 nm range.}
\label{FIG:Shalaev}
\begin{center}
    \includegraphics[width=7cm]{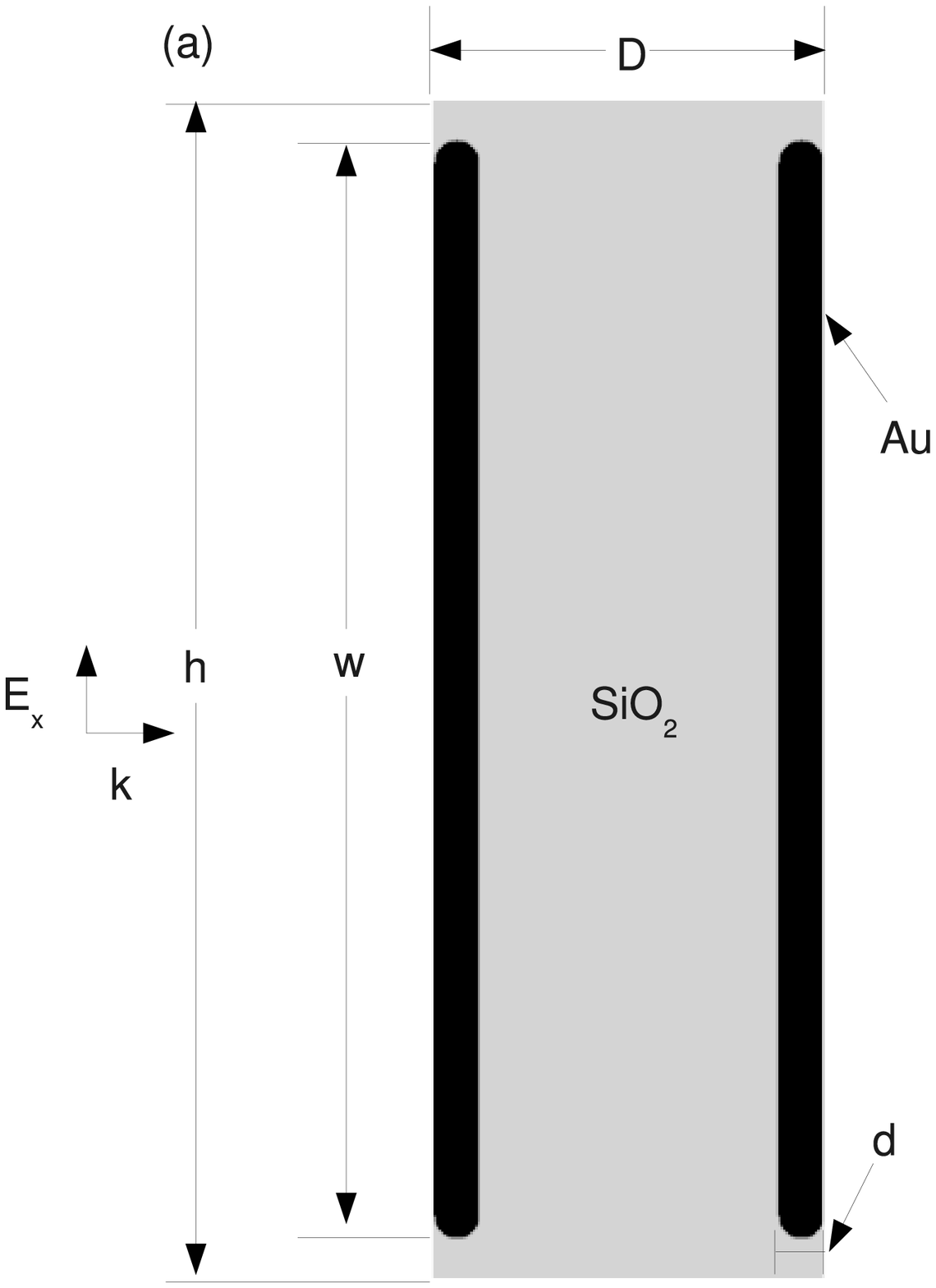}
    \hspace{-2cm}
    \includegraphics[width=6cm]{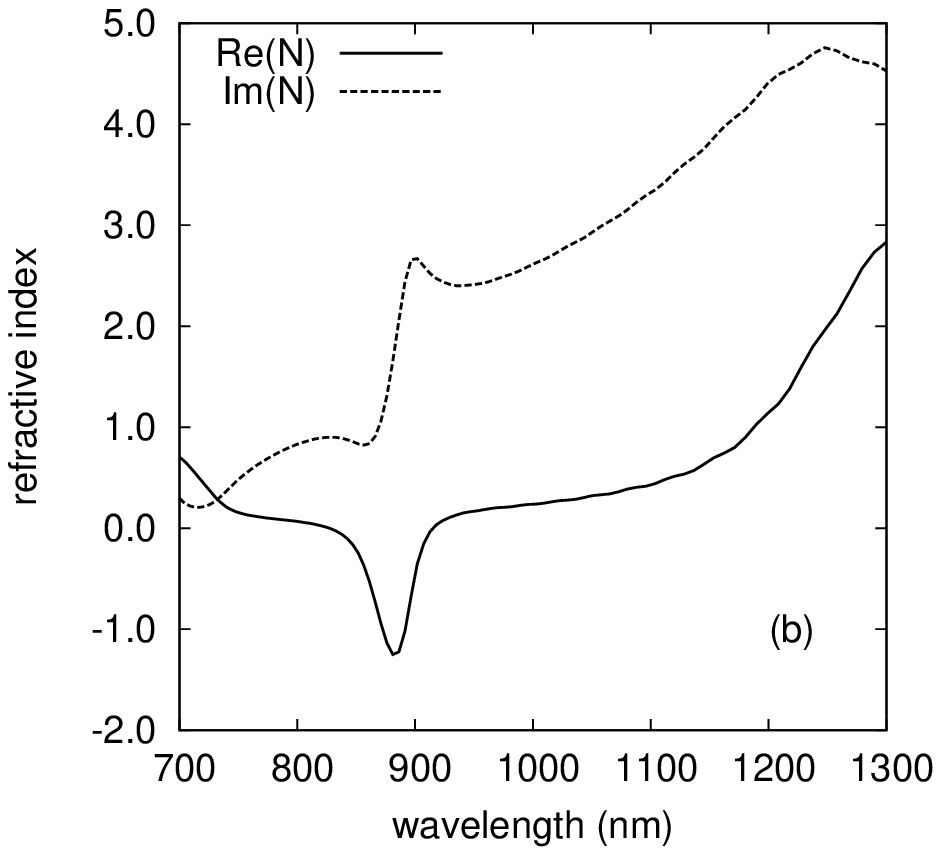}
\end{center} 
\end{figure}

\begin{figure}[!htpb]
\caption {Real (a) and imaginary (b) components of the refractive index 
extracted according to Eq. (\ref{EQN:REFIND}) as well as transmission 
(c) and reflection (d) spectra for various rod thicknesses, d.  All other 
parameters are held constant at the values detailed in the text.}
\label{FIG:vary_d}
\begin{center}
    \includegraphics[width=6.5cm]{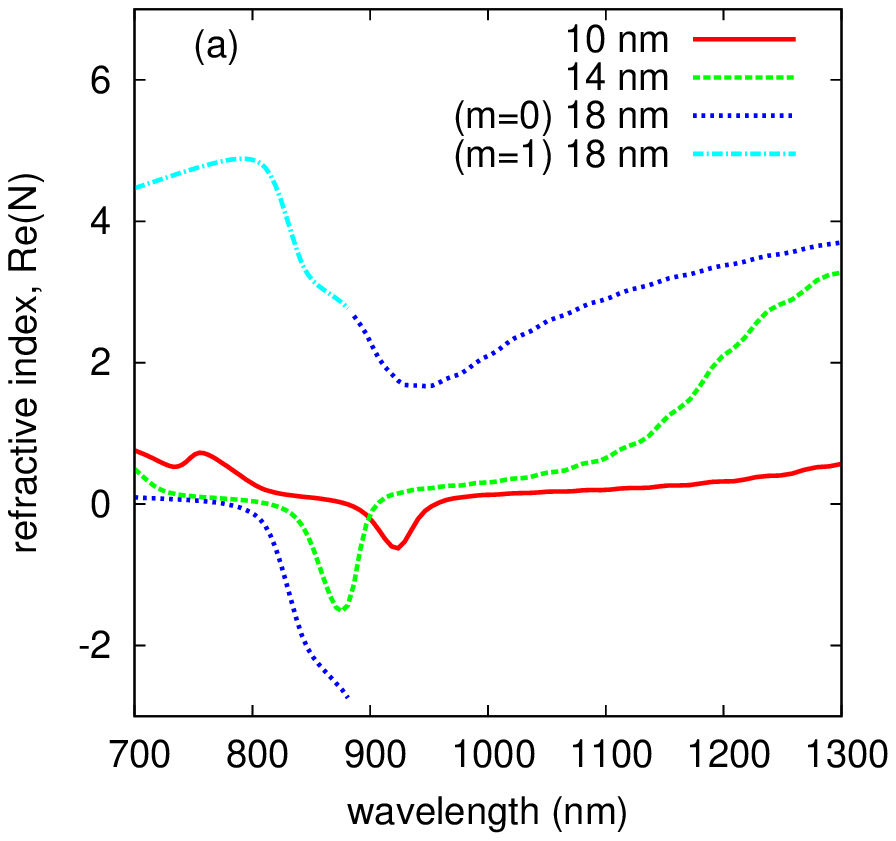}
    \includegraphics[width=6.5cm]{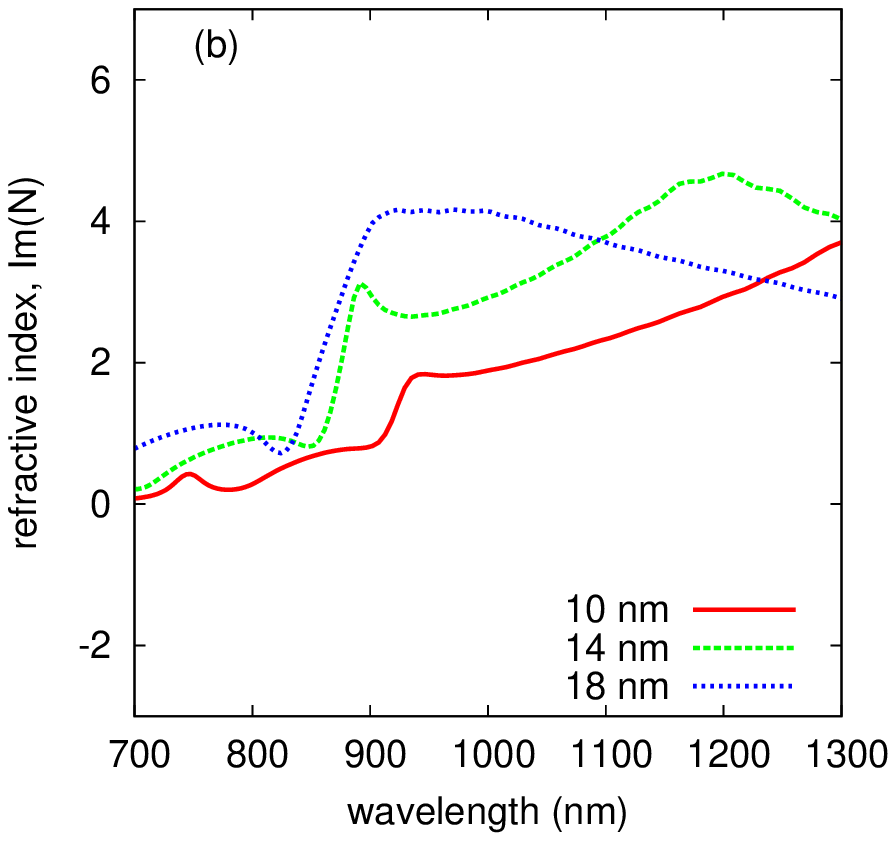}
    \includegraphics[width=6.5cm]{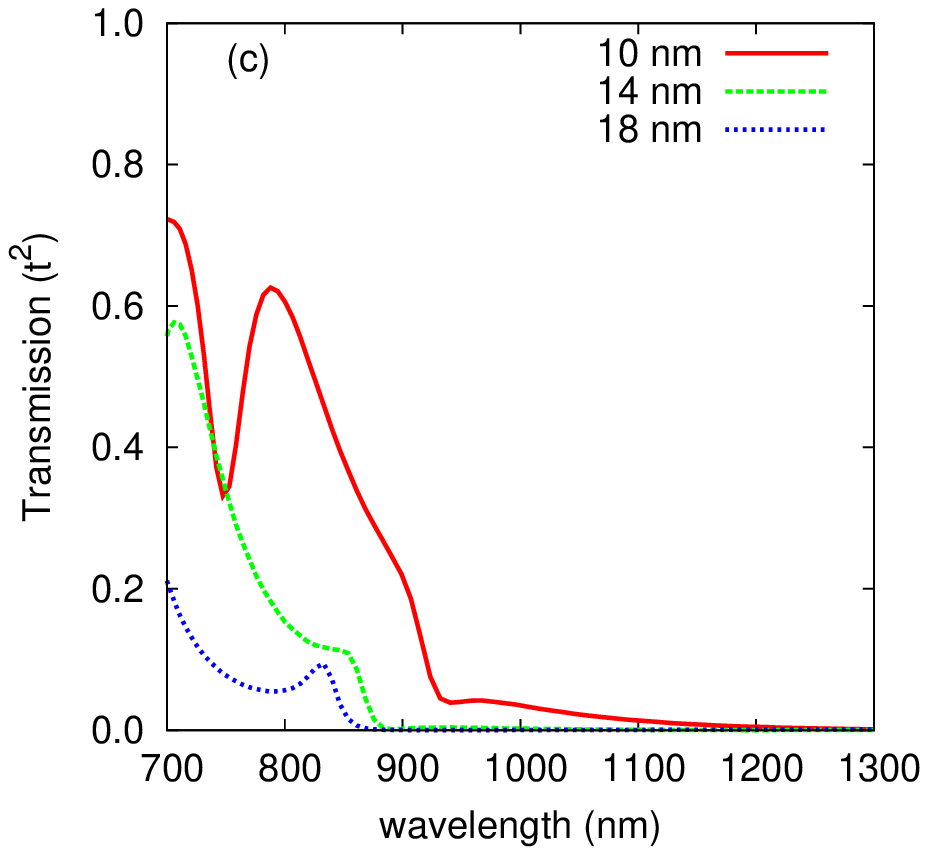}
    \includegraphics[width=6.5cm]{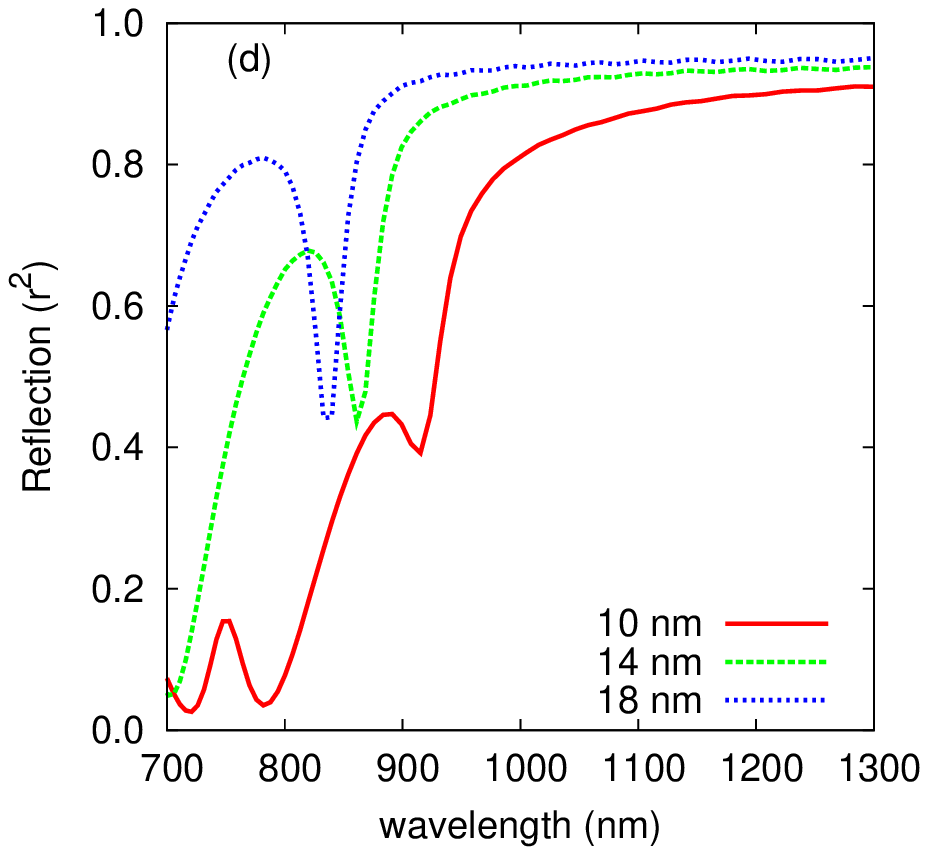}
\end{center} 
\end{figure}

We consider a two-dimensional system of gold nanorods separated by SiO$_2$ 
similar to that  investigated by Shalaev {\em et al.} \cite{REF:Shalaev06-3}.  
The system is illustrated in Fig. \ref{FIG:Shalaev}(a), with the parameters 
given in Ref. \cite{REF:Shalaev06-3} to obtain negative refractive index; the 
rods of thickness d=13 nm and width w=450 nm are separated by SiO$_2$ with a
total slab thickness of D=160 nm.  The system is periodic in the x-direction
with a periodicity of h=480 nm.  We obtain the complex transmission and 
reflection coefficients with finite-difference time-domain (FDTD) simulations as
implemented in the freely available Meep FDTD package \cite{REF:MEEP}. This 
particular structure, as noted in Ref \cite{REF:Shalaev06-3} and depicted in
Fig \ref{FIG:Shalaev}(b), clearly exhibits a negative refractive index in the 
850-900 nm region.  
Our present results reproduce well those given in Ref. \cite{REF:Shalaev06-3}.
As an aside, we note that the permeability for this structure is positive; 
the structure is a NIM but {\em not} a double negative material.  The 
material does, however, satisfy the necessary conditions given in Ref. 
\cite{REF:Depine04}.  
Optimizing this structure for negative real component of the refractive index,  
we consider first only varying the rod thickness, d, while holding all 
parameters at their values listed above.  Figure \ref{FIG:vary_d} illustrates 
the real and imaginary components of the refractive index obtained from 
the extraction in Eq. (\ref{EQN:REFIND}) as well as transmission and 
reflection spectra.  The real component of the refractive index for the d = 10 
and d = 14 nm cases (Fig. \ref{FIG:vary_d}(a)), like the d = 13 nm case, is 
unambiguously determined by the extraction procedure.  However, a discontinuity 
develops in the $m$=0 branch for d = 18 nm.  As stated above, 
the macroscopic refractive index should be independent of slab thickness, and
this property should be exploited in assigning branches.  In practice,
though, a NIM almost never exhibits this nice property, and the assignment must
be made on the basis of maintaining the continuity of the refractive index as
a function of frequency. A continuous refractive index function can be obtained
for d = 18 nm by combining the $m$=0 and $m$=1 branches, resulting in the
upper continuous curve. However, this choice gives the refractive index as a 
discontinuous 
function of the rod thickness, d.  One would expect that such an abrupt change 
in character of the refractive index would be associated with an equally abrupt 
change in either the imaginary component of the refractive index or the 
transmission or reflection spectra, but each of these do indeed vary smoothly 
with the geometric parameter, d.  The imaginary component of the refractive 
index does lose a sharp peak when increasing d from 14 to 18 nm, but this 
change is subtle compared to that suggested by Fig. \ref{FIG:vary_d}(a).  We
propose that the correct form of Re(N) may be obtained unambiguously by the
integration of Im(N) according to Eq. (\ref{EQN:KK_real}).

\begin{figure}[!htpb]
\caption {Real and imaginary components of the refractive index as obtained from the extraction in Eq. (\ref{EQN:REFIND}) and, for the real component, the 
  Kramers-Kronig integration of Eq. (\ref{EQN:KK_real}).  Results are presented
  for a gold nanorod system with d=13 nm.}
\label{FIG:KK_13_160}
\begin{center}
    \includegraphics[width=6.5cm]{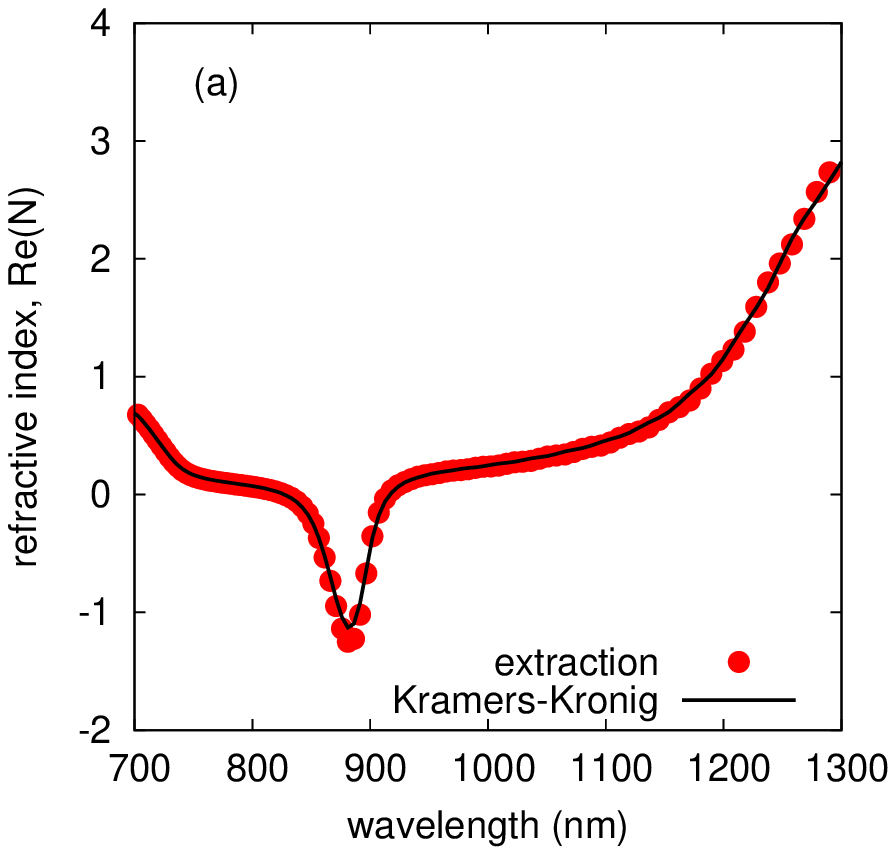}
    \includegraphics[width=6.5cm]{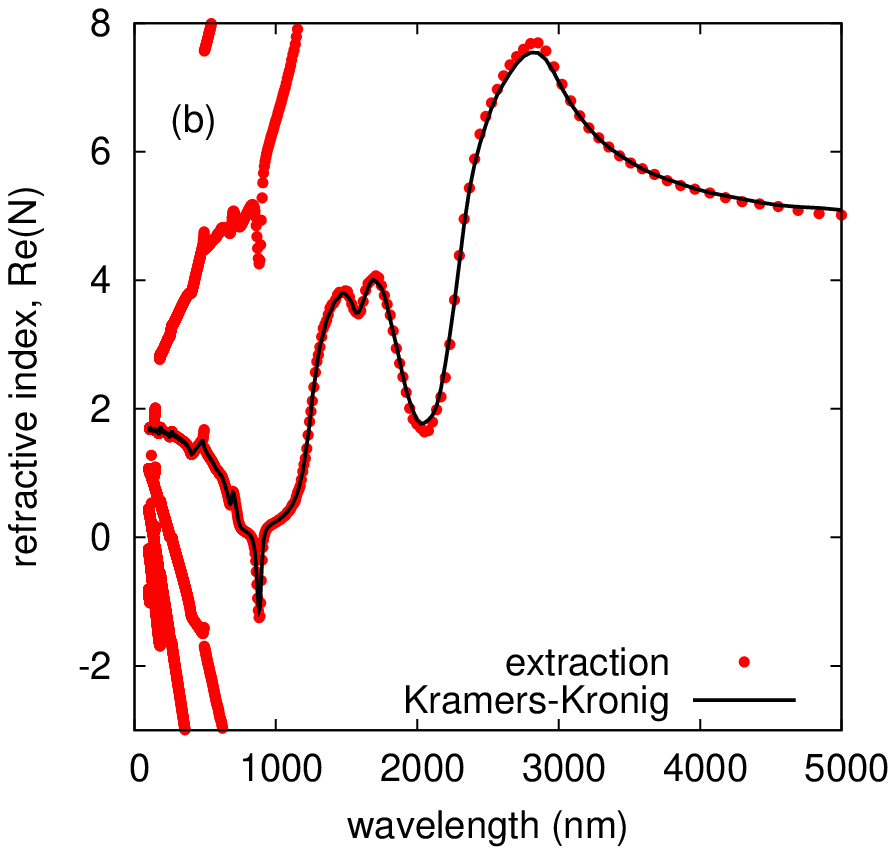}
    \includegraphics[width=6.5cm]{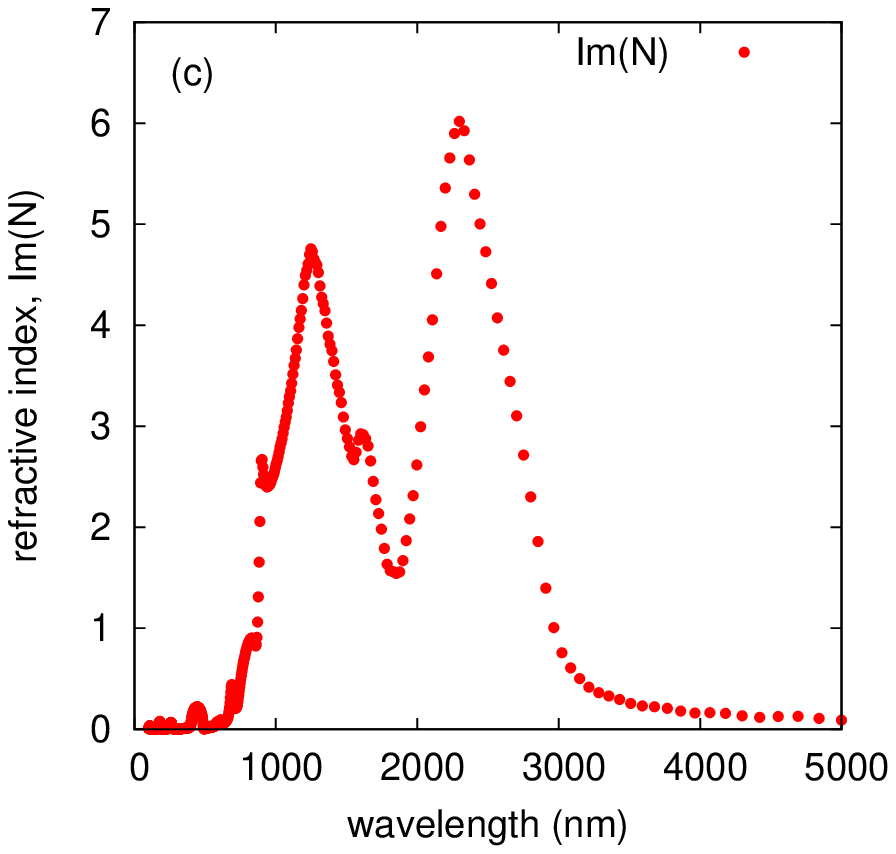}
\end{center} 
\end{figure}

Consider again the well-behaved system (d=13 nm) in which negative
refractive index is observed in the 850-900 nm range.  The values for the index
extracted from Eq. (\ref{EQN:REFIND}) and those obtained by integration of Eq.
(\ref{EQN:KK_real}) should be equivalent, provided that we have integrated Im(N)
over an appropriately large range of frequencies.  Table \ref{TAB:KK_INT} 
lists the non-parallelity errors (NPE) obtained for Re(N) obtained from the KK 
integration as compared to that obtained from the extraction procedure for 
several different wavelength ranges.
\begin{table}[!htpb]
  \caption{Non-parallelity error (NPE) in the Kramers-Kronig integration as compared to extracted values of Re(N). The NPE is defined as the difference between the most positive and most negative errors between the KK and extracted values of Re(N) over the range 700-1300 nm.}
  \label{TAB:KK_INT}
  \begin{center}
    \begin{tabular}{ccc}
    \multicolumn{2}{c}{wavelength (nm)} \\
    \cline{1-2}
    minimum & maximum & NPE \\
    \hline
    \hline
    700 & 1300 & 5.434 \\
    700 & 1500 & 1.224 \\
    700 & 2000 & 0.742 \\
    700 & 5000 & 0.478 \\
    100 & 1300 & 5.071 \\
    100 & 1500 & 0.862 \\
    100 & 2000 & 0.380 \\
    100 & 5000 & 0.200 \\
    \hline
    \hline
    \end{tabular}
  \end{center}
\end{table}
NPE are defined here by taking the difference between the most positive and most
negative errors in the KK integration as compared to the extracted values of 
Re(N) in the range of 700-1300 nm. Integration over only the range of interest 
(700-1300 nm) leads to very large errors, and the integrated values of 
Re(N) are virtually meaningless.  The KK value for Re(N) become quite accurate
when we extend the integration range to 100-5000 nm, as is evidenced by the NPE
in Table \ref{TAB:KK_INT} and visually in Figs. \ref{FIG:KK_13_160}(a) and (b).
The KK integration reproduces the negative region of refractive index in the 
850-900 nm region as well as all other features from 100-5000 nm.  
The very large and very wide features present in Im(N) in Fig. 
\ref{FIG:KK_13_160}(c) illustrate the importance of such a wide range of 
frequencies for obtaining reliable results; these broad features significantly 
contribute to the KK integral. 

\begin{figure}[!htpb]
\caption {Real component of the refractive index as obtained from the extraction
 in Eq. (\ref{EQN:REFIND}) or the Kramers-Kronig integration of Eq. 
(\ref{EQN:KK_real}) for a gold nanorod system with d=18 nm.}
\label{FIG:KK_18_160}
\begin{center}
    \includegraphics[width=6.5cm]{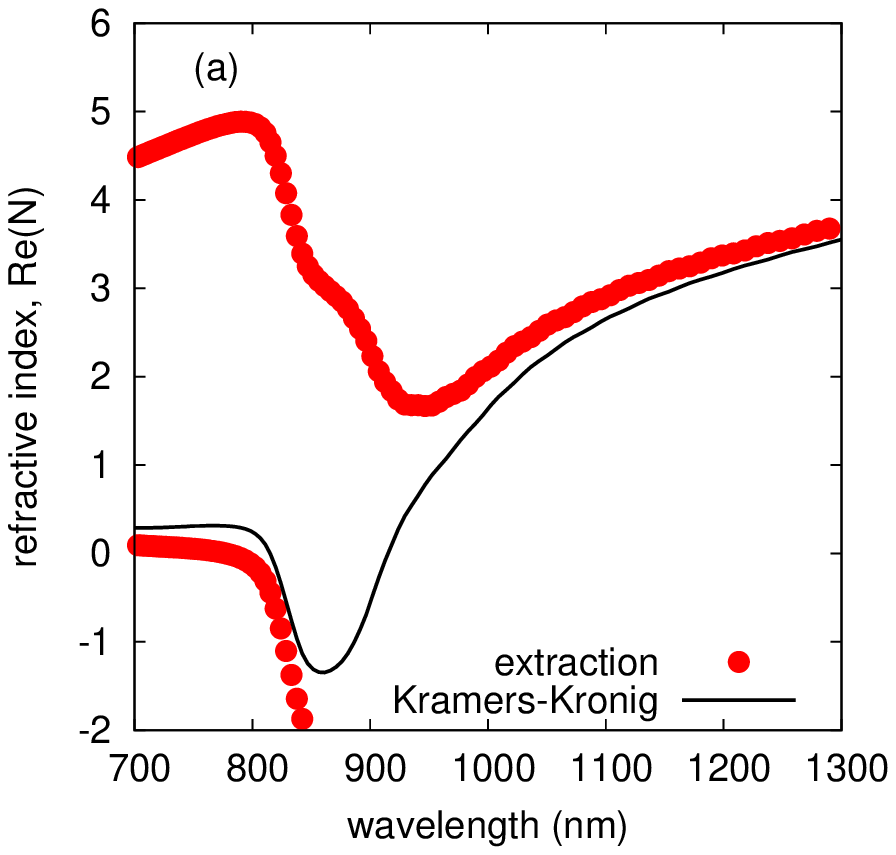}
    \includegraphics[width=6.5cm]{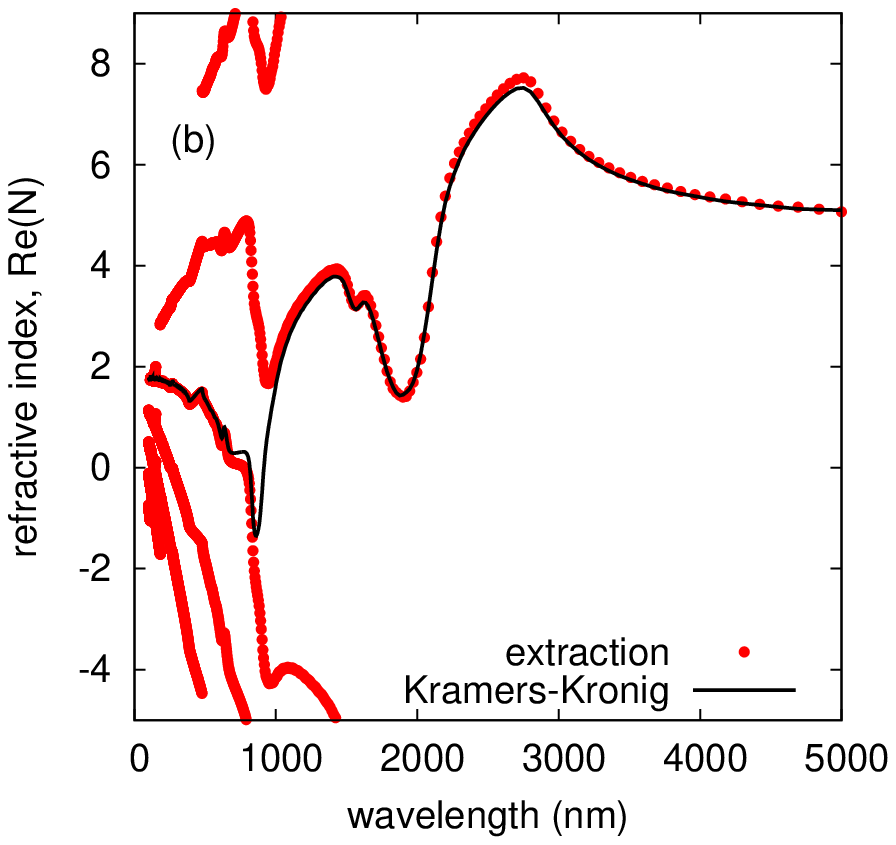}
\end{center} 
\end{figure}

Now consider the d=18 nm case in which a discontinuity developed 
in the $m$=0 branch, and the choice for the correct branch is 
made subjectively.  Forcing Re(N) to be a continuous function of wavelength 
would require one to combine the $m$=0 and $m$=1 branches to form the upper 
curve in Fig. \ref{FIG:vary_d}(a), but as shown in Fig. \ref{FIG:KK_18_160}, 
the correct choice as determined by KK integration involves {\em only} the
$m$=0 branch in the 700-1300 nm range, yielding a negative index with a 
minimum around 880 nm.
The original branch assignment based solely on the extraction procedure and
continuity of the refractive index predicts no negative index and is thus 
qualitatively incorrect.  Ambiguous situations such as this where index 
assignments are based on the continuity of Re(N) are surprisingly common, and 
the KK integration can provide a very easy and necessary check as to the 
validity of branch assignments.  The question remains as to {\em why} the
extraction procedure results in a spurious discontinuity in such an important
frequency range. In general we observe that, for a gold nanorod NIM of fixed 
width, D, as the ratio of d to D increases, a false discontinuity inevitably 
develops in the extracted values of Re(N).  The KK integration, by enforcing
causality, can be used to interpolate through these problematic regions and
select the proper branches on either side.

\begin{figure}[!htpb]
\caption {Real component of the refractive index as obtained from the extraction
 in Eq. (\ref{EQN:REFIND}) or the Kramers-Kronig integration of Eq. 
(\ref{EQN:KK_real}) for a gold nanorod system with (a) FOM=1.517 and (b) 
FOM=1.442.}
\label{FIG:KK_phi}
\begin{center}
    \includegraphics[width=6.5cm]{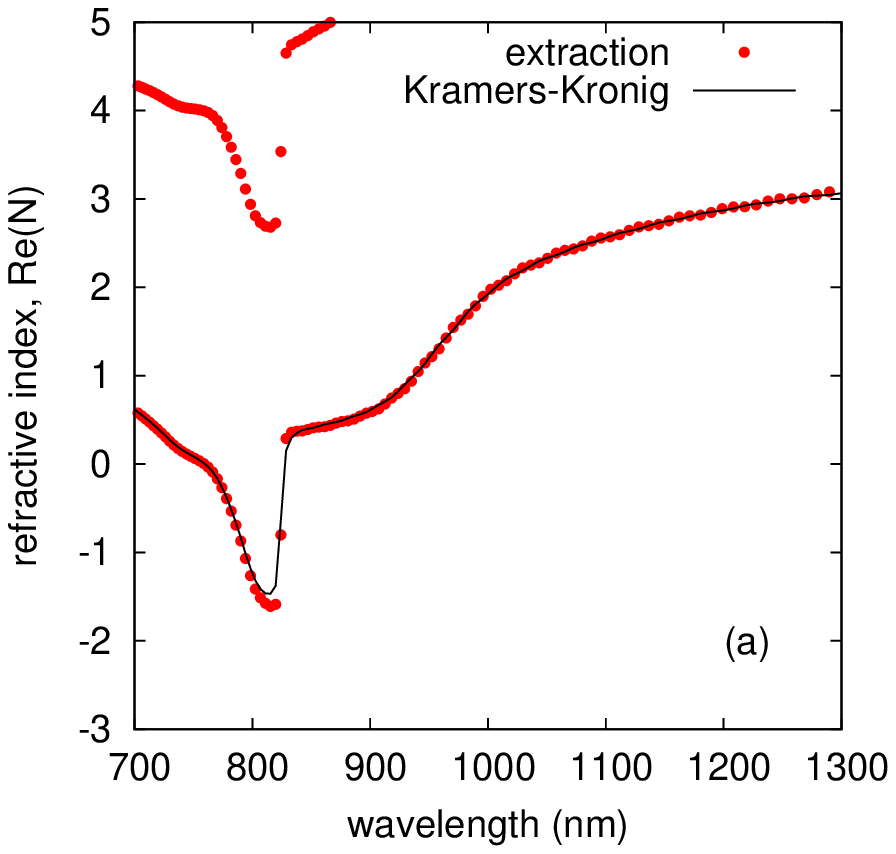}
    \includegraphics[width=6.5cm]{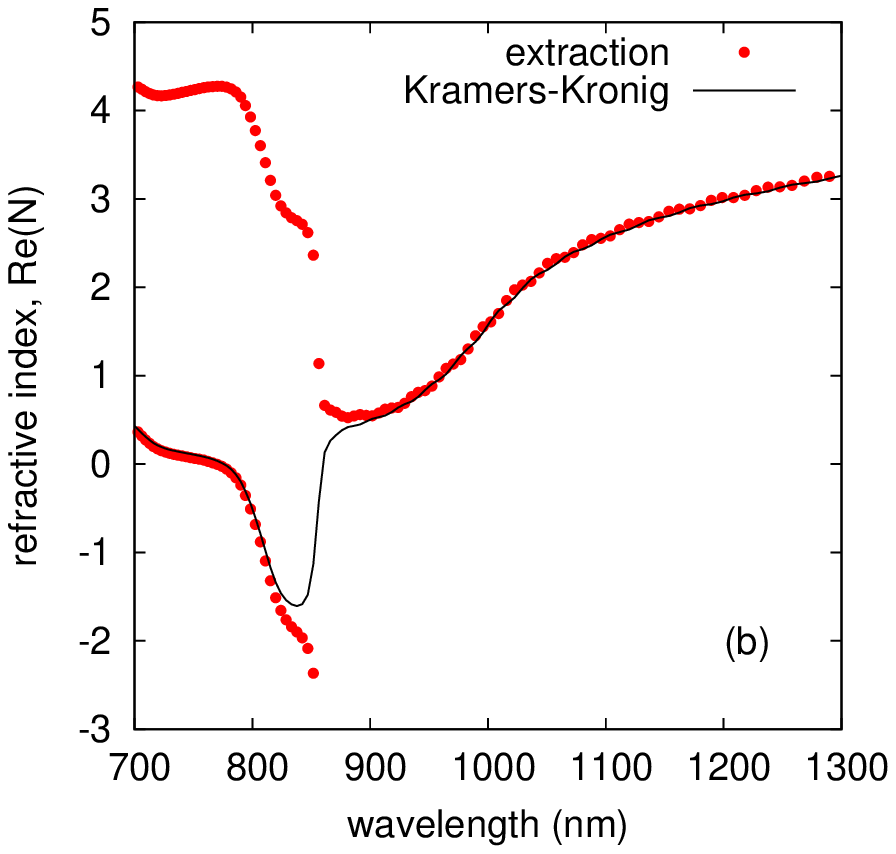}
\end{center} 
\end{figure}

The combined extraction/KK procedure for determining refractive indices, in
removing ambiguities in branch assignments, allows for a more ``black-box" 
approach to the optimization of NIM for either very negative refractive 
index or for a NIM quality factor or figure of merit (FOM) related to the 
ratios of the real and imaginary components if N:
\begin{equation}
\label{EQN:RATIO}
{\rm FOM} = - \frac{Re(N)}{Im(N)}.
\end{equation}
We have optimized the structure in Fig. \ref{FIG:Shalaev}(a) for the FOM by 
varying 
the parameters d, D, and w, sampling a total of 768 structures with d, D, and w 
in the ranges 12-19 nm, 120-230 nm, and 400-470 nm, respectively.  The structure 
exhibiting the largest FOM as given by Eq. (\ref{EQN:RATIO}) was
found to have d=17 nm, D=190 nm, and w=430 nm, with a NIM figure of merit,
FOM = 1.517.  This value is nearly double the value of 0.829 that corresponds
to the structure suggested in Ref. \cite{REF:Shalaev06-3} with d=13, D=160, and
w=450 nm, and the optimized material is a double negative material, where
both $\mu$ and $\epsilon$ are simultaneously negative.  Recall that the 
FOM=0.829 structure of Ref. \cite{REF:Shalaev06-3} has a positive relative 
permeability.  Fourteen structures were 
found to have figures of merit greater than 1.4, and, surprisingly, four of 
these 
structures might not have been found with the simplistic, continuous-function 
approach to the extraction procedure.   Figure \ref{FIG:KK_phi} depicts the 
$m$=0 and $m$=1 brances of Re(N) for the optimal FOM=1.517 as well as the best 
structure found that exhibits a false discontinuity in the extracted index, with 
FOM=1.442 (d = 17 nm, D = 180 nm, w = 440 nm).  
Clearly, the KK relations allow us to easily and unambiguously select the 
proper branches of Re(N) obtained from the retreival.  Amazingly, the 
high-quality structure whose refractive index is represented in Fig. 
\ref{FIG:KK_phi}(b) would not have been considered a NIM without the KK 
analysis. This FOM=1.442 structure, like the optimal FOM=1.517 structure, is 
also a double negative material.

\section{Conclusions}
We have developed a reliable and robust methodology for extracting effective
refractive indices for metamaterials by combining the usual S-parameter 
extraction procedures with the concept of causality as expressed in the 
Kramers-Kronig relationship. Causality 
absolutely must be enforced when extracting effective parameters from 
the three-layer homogenous metamaterial model.  Our results suggest that 
simplistic extraction models in which one copes with branching ambiguities by
enforcing continuity of the refractive index as a function of frequency may 
result in  qualitatively incorrect results.  In addition, in regions where Im(N)
is very close to zero, the sign chosen for Eq. (\ref{EQN:REFIND}) may suffer 
from numerical noise; this sign issue has no implications for Im(N), as it is 
virtually zero, but Re(N) may be given the incorrect sign.  Re(N) obtained from 
KK integration will be insensitive to this issue when Im(N) is close to zero
and will thus yield more reliable results. The Kramers-Kronig relations 
naturally guide branch selection and are simple enough that they may be used in
a ``black-box" sense for the optimization of metamaterial geometries for 
desired properties.  The present methodology was used to optimize a gold 
nanorod system for a NIM quality factor, $Q$, resulting in a structure with a 
$Q$-factor nearly double that of the original starting structure.  Several
high-quality structures were also obtained that would not have been considered
NIM with existing extraction procedures.

The KK relations can also provide a useful gauge as to the validity of the
assumption of metamaterial homogeneity that is central to the S-parameter
retrieval procedure.  One can view the false discontinuities in the refractive
index functions as a the first symptoms of the breakdown of the validity of the 
homogeneous three-layer model. In extreme cases, where the amount of metal 
relative to the dielectric present in the sample becomes significant, the KK 
relation of Eq. (\ref{EQN:KK_real}) can only qualitatively reproduce the 
extracted index over the entire frequency spectrum.  The KK results are not 
incorrect; the discrepency merely reflects the failure of the assumption of the
homogeneous nature of the metamaterial.  In these rare instances, the 
extraction procedure as a whole may not be an appropriate way to obtain 
refractive indices. 

\section{Acknowledgments} 
This research used resources of the Argonne Leadership Computing Facility and the Center for Nanoscale Materials at Argonne National Laboratory, which are supported by the Office of Science of the U.S
. Department of Energy under contract DE-AC02-06CH11357.
AED thanks Jeff Hammond for support on the BlueGene/P system in the Argonne Leadership Computing Facility. We thank Matt Pelton for
helpful comments.  AED acknowledges funding from the Computational Postdoctoral fellowship through the Computing, Engineering, and Life Sciences Division of Argonne National Laboratory.
\end{document}